# Wavelet Selection and Employment for Side-Channel Disassembly

Random Gwinn
*Johns Hopkins University*
*Applied Physics Laboratory*
Laurel, MD, USA
random.gwinn@jhuapl.edu

Mark Matties
*Johns Hopkins University*
*Applied Physics Laboratory*
Laurel, MD, USA
mark.matties@jhuapl.edu

Aviel D. Rubin
*Johns Hopkins University*
*Whiting School of Engineering*
Baltimore, MD, USA
rubin@jhu.edu

*Abstract*— Side-channel analysis, originally used in cryptanalysis is growing in use cases, both offensive and defensive. Wavelet analysis is a commonly employed time-frequency analysis technique used across disciplines, with a variety of purposes, and has shown increasing prevalence within side-channel literature.

This paper explores wavelet selection and analysis parameters for use in side-channel analysis, particularly power side-channel-based instruction disassembly and classification. Experiments are conducted on an ATmega328P microcontroller and a subset of the AVR instruction set. Classification performance is evaluated with a time-series convolutional neural network (CNN) at clock-cycle fidelity. This work demonstrates that wavelet selection and employment parameters have meaningful impact on analysis outcomes. Practitioners should make informed decisions and consider optimizing these factors similarly to machine learning architecture and hyperparameters. We conclude that the *gaus1* wavelet with scales 1-21 and grayscale colormap provided the best balance of classification performance, time, and memory efficiency in our application.

*Keywords*— *side-channel analysis, wavelet transform, scalogram, disassembly, classification*

I. INTRODUCTION

Side-channel analysis gained significant attention and adoption for use in cryptanalysis after Kocher's seminal work [1]-[2] in the late 1990s. Since then, additional side-channels such as: electromagnetic [3], thermal [4], optical [5], methods such as template [6], and use cases such as detecting malware [7] and hardware trojans [8]-[9], reverse engineering [10], and disassembly [11]-[15] have been increasingly explored.

Wavelet analysis has been leveraged for myriad purposes, such as signal filtering [16]-[18], data compression [18]-[20], and feature analysis or extraction. As a feature analysis and extraction technique its vast usage examples include electrocardiogram [21], seismic data [22], 3D spatial reconstruction [23], and side-channel analysis [14], [18], just to name several. Prior work exists towards establishing wavelet selection methods, but these are largely domain and application specific [23]-[26].

As above establishes, there exist numerous uses for both side-channel analysis, and wavelet analysis, including multiple purposes for the use of wavelet analysis in support of side-channel analysis. However, the authors of this work are not aware of prior work that deliberately explored and described wavelet selection methods or employment considerations for side-channel analysis. Further, existing side-channel literature leveraging wavelet analysis methods are largely silent on specific implementations, which can lead to issues with reproducibility. Additionally, as some factors include trade-offs, this is important to explore in order to bring techniques out of the research laboratory and into engineering practice.

Contributions of this work include but are not necessarily limited to: first known exploration of wavelet selection and employment considerations for side-channel analysis, quantified by classification performance; establishing side-channel measurements as a function of the ideal signal and associated noise, measurement uncertainty, and variance (6); assessment of a wavelet selection technique from non-side-channel literature for use in side-channel disassembly, with comparison to actual results.

This paper is organized with background of time-frequency analysis and side-channel analysis in section II; an overview of the evaluation approach, including experimental setup, wavelets explored, and wavelet selection considerations in section III; experimental results in section IV; and conclusions in section V.

II. BACKGROUND

The following subsections provide a high-level overview and introductory context for the time-frequency analysis and side-channel analysis techniques employed within this paper.

## A. Time-Frequency Analysis

Several techniques exist within the area of time-frequency analysis which explores input signals in both the time and frequency domain concurrently to garner additional insights than inspecting each domain separately. Two common techniques are the Short-Time Fourier Transform (STFT) and wavelet transforms. Wavelet transform analysis offers advantages over the more traditional STFT as discussed in the following subsections.

### 1) Fourier Transform

Whereas the Fourier series approximates periodic functions through sums of harmonically-related sine and cosine functions [27], the Fourier transform provides a similar capability that can be applied to aperiodic functions [27]. For our purposes, this enables the analysis of time series signals through their approximated frequency components. However, it does not provide any temporal localization of those components. A standard representation of the Fourier transform [28] is shown in (1). Fig. 1 provides examples of the Fast Fourier Transform (FFT) (a.2, b.2) of side-channel input signals (a.1, b.1).

$$(Ff)(\omega) = \frac{1}{\sqrt{2\pi}} \int dt\, e^{-i\omega t}\, f(t) \qquad (1)$$

### 2) Short-Time Fourier Transform

The STFT computes the Fourier transform in windowed-fashion, $g$, along the length of an input signal or function $f$. This provides some degree of temporal localization of spectral components, such that the STFT, or $(T^{win}f)(\omega,t)$, "can be interpreted loosely as the 'content' of $f$ near time $t$ and near frequency $\omega$ [28]." The discrete version of the STFT is shown in (2), with $t$ assigned interval $nt_0$, and $\omega$ assigned interval $m\omega_0$. While this provides a combined temporal-frequency view, due to inherent uncertainty the resolution is limited by window-size and involves tradeoffs between domains. With narrow windows, temporal resolution is increased while frequency resolution decreases; whereas with broad windows the opposite is true.

Fig. 1 provides examples of the STFT depicted as spectrograms (a.3, b.3) of side-channel input signals (a.1, b.1). The STFT limitations, are observable in both (a.3) and (b.3) through blocky-representations.

$$T^{win}_{m,n}(f) = \int ds\, f(s)\, g(s - nt_0)e^{-im\omega_0 s} \qquad (2)$$

### 3) Wavelet Transform

Wavelet transforms overcome the STFT resolution tradeoff by executing multiresolution rather than fixed analysis. Like the STFT, wavelet transforms can be applied with continuous or discrete variants. However, instead of calculating the Fourier transform of fixed windows, a wavelet function is variably scaled and translated over the input signal producing coefficients which are combined as the multiresolution time-frequency output. This enables superior ability of the wavelet transform over the STFT "to 'zoom in' on very short-lived high frequency phenomena" such as transients or singularities [28]. Equations (3) and (4) describe the continuous wavelet transform function [28].

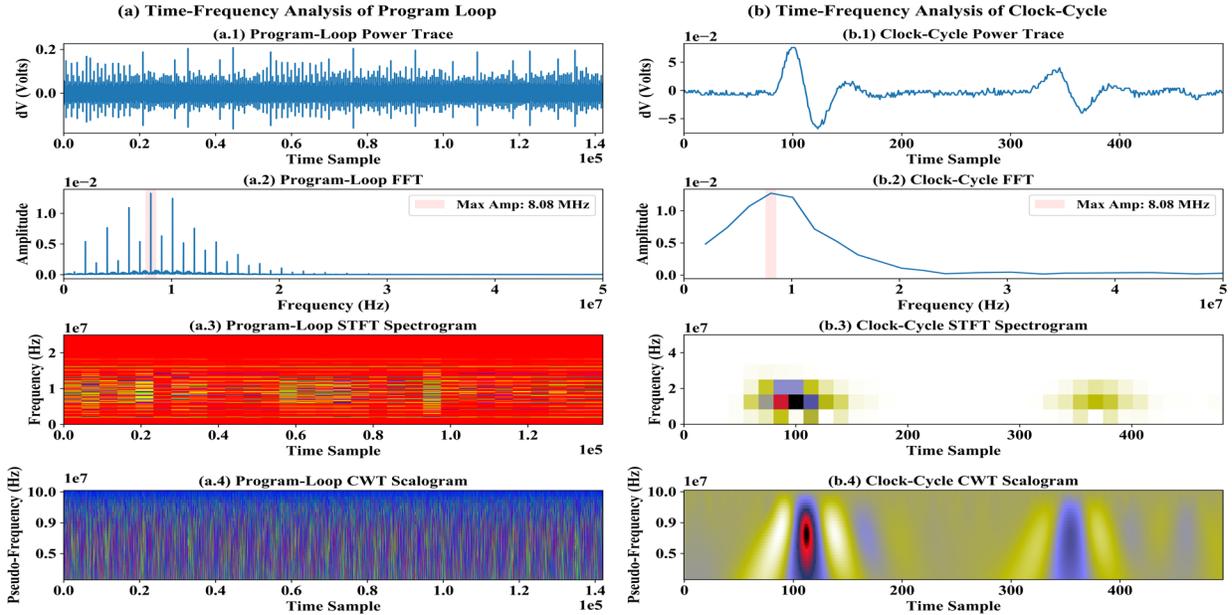

Fig. 1: Example time-frequency analysis of program loop (a) and clock-cycle (b) each with raw trace (1), FFT (2), STFT (3), and Continuous Wavelet Transform (CWT) (4). The program-loop STFT (a.3) uses a window size of 5,000 samples with 10% overlap and *prism* colormap while the clock-cycle STFT (b.3) uses a window size of 60 samples with 80% overlap and *gist_stern_r* colormap. The CWT examples use the *gaus1* wavelet with coefficients calculated for scales 1-29, with *prism* colormap for the program-loop (a.4) and *gist_stern_r* colormap for the clock-cycle (b.4). The *prism* colormap was employed for the program-loops due to its ability to provide visual distinction at high sample densities.

Fig. 1 provides examples of the continuous wavelet transform depicted as scalograms (a.4, b.4) of side-channel input signals (a.1, b.1). The y-axis was converted to show pseudo-frequency instead of scale for more ready comparison to the STFTs.

$$(T^{wav}f)(a,b) = |a|^{-1/2} \int dt\, f(t)\, \psi\left(\frac{t-b}{a}\right), \quad (3)$$

with $\psi$ satisfying:

$$\int dt\, \psi(t) = 0. \quad (4)$$

A baseline wavelet function, of which many exist, is often referred to as a "mother wavelet", and is represented by $\psi$. Specific instances of these functions with scale $a$ and translation factor $b$, are represented by $\psi^{a,b}$, and referred to as "wavelets." [28] Frequency and scale have an inverse relationship in wavelet transforms, with small scales corresponding to high frequencies and large scales corresponding to low frequencies. Additionally, the scale and frequency correspondence is imprecise, so approximations, or pseudo-frequencies are referenced instead by [29]:

$$F_a = \frac{F_c}{a \times \Delta}. \quad (5)$$

The pseudo-frequency, in Hz, at a given scale $a$, is $F_a$, while $F_c$ is the wavelet's center frequency in Hz, and $\Delta$ represents the sampling period in seconds.

### B. Side-Channel Analysis

Side-channels, generically, are unintended sources of information. For computing systems this usually refers to sensitive information about the system and its internal operations that can be gleaned through monitoring and analysis of one or more of the system's side-channels. Numerous side-channels exist, with power [2] and electromagnetic emissions [3] being especially common.

Typically, side-channel analysis involves extracting cryptographic keys or other private information, and some level of access to the system is assumed: direct for power or proximity (usually near field) for electromagnetic emissions. For power side-channels, simple or differential power analysis [2], template [6], or other analysis means would then be employed to extract the desired information from the captured side-channel trace data.

In this work, the use case of disassembly through power side-channels is explored. This follows closely to other analysis methods where large amounts of side-channel data are collected and analyzed ahead of time, but instead of looking for a specific correlation (e.g., the correct cryptographic key in the haystack of trace data), all trace information is intended for correlation to a corresponding system state and function, namely the set and sequence of executed processor instructions. This information can then support a variety of end purposes, such as reverse engineering or integrity monitoring of the device under test (DUT).

Noise and uncertainty are facts of life within side-channel measurement and analysis. These factors are introduced by a variety of sources such as noisy power input, electromagnetic noise, uncertainty in measurements of even calibrated equipment, as well as variability in the device(s) under measurement. A generic representation of these factors, introduced in this work, is shown as:

$$T_{meas} = S + (N_1 + \ldots N_n) + (M_1 + \ldots M_m) + (V_1 + \ldots V_v) \quad (6)$$

where $T_{meas}$ is the measured side-channel trace, $S$ is the ideal signal, $N$ represents individual noise sources, $M$ represents individual measurement uncertainties, and $V$ represents the variability in the device(s) under measurement, to include non-noise induced environmental factors. Accounting for factors in (6) can be key in achieving the objectives for a given side-channel use case effectively and efficiently. This can also support a more robust solution, as the variability between devices under measurement and within a given device over its operational life are important considerations.

Examples of $T_{meas}$ are shown in Fig. 1 (a.1, b.1), with the impact of $N$ being particularly apparent as serrated features along the main signal pattern in the clock-cycle trace of (b.1), which largely disappear when many traces of the same signal are averaged.

### III. EVALUATION APPROACH

This section is intended to baseline the reader on the basic experimental setup along with the scope and methods of wavelet evaluation within this work. As the intent of this work is to highlight the impact and considerations for wavelet employment, a detailed description of the classification approach is not included. Additionally, a subset of the AVR instruction set [30] was selected across instruction types and clock-lengths to establish general impacts and considerations towards the central purpose of wavelet employment. The authors would point readers towards the large body of work in either area: classification and side-channel disassembly, for additional reading on the state of the art in each.

*A. Experimental Setup and Baseline Analysis*

An Elegoo Uno R3 served as the DUT, with an ATmega328P microcontroller down-clocked to 1 MHz. A digital sampling oscilloscope with a sampling rate of 500 MSa/s and bandwidth of 200 MHz was used to capture the power side-channel traces from voltage drop across a 1 kΩ shunt-resistor on the ground-side.

The DUT executed a program loop with each instruction from the included subset (listed in Table I) paired with an instance of every other subset instruction. The exception being the *rjmp* instruction was only used to restart the looped instruction sequence. The program loop consisted of 191 instructions over 287 total clock-cycles of execution.

Five separate side-channel trace files were collected at a memory depth of 14 Mpts. This provides for 485 complete program loops, and 9,215 instances of each instruction (except *rjmp*), for analysis.

Analysis was performed in Python 3 notebooks with the PyWavelets [31], NumPy [32], and SciPy [33] packages for time-frequency analysis and TensorFlow [34] with Keras [35] for classification. A time-series CNN, with wavelet coefficients of each clock-cycle window transformed into scalograms was used for classification, with a 70/30 train/test split of data. A semi-optimized set of scalogram, model architecture, and hyper-parameters was selected as the baseline for wavelet transform assessment that provided a balance between classification performance and time complexity, but also helped avoid overfitting. Specific optimizations are considered outside the scope of this paper.

Since the use-case explored within this work is side-channel disassembly, the evaluation metrics employed are instruction classification accuracy and computational time. Scalogram and CNN memory size, relative to color-depth, is also considered. Conventional cryptanalytic side-channel metrics, such as success rate, guessing entropy, and key rank are not relevant here.

*B. Wavelets Examined*

Within the PyWavelets package, wavelets are grouped into 14 families, each containing one or more mother wavelets. In total, there are more than 120 built-in mother wavelets, as well as the ability to introduce custom mother wavelets. Mother wavelets are further categorized as discrete or continuous, by conditions such as orthogonality, symmetry, real or complex, and by attributes such as lower and upper bounds, bandwidth frequency, and center frequency.

This work explores the continuous and real mother wavelets listed in Table II and depicted in Fig 2. Additionally, of the wavelets explored, none were orthogonal or bi-orthogonal, so that condition was excluded from the table. Furthermore, as only the complex wavelet families Complex Morlet, Shannon, and Frequency B-Spline include a bandwidth parameter, it is also excluded from Table II. The upper and lower bounds of the included mother wavelets are shown as the x-axis bounds in Fig. 2.

The mother wavelets in the Gaussian family, *gausN$_{ord}$*, represent derivatives of the Gaussian function [36]:

$$\text{Gaus}(x) = \frac{1}{\sigma(\sqrt{2\pi})} e^{-(x-\mu)^2 / (2\sigma^2)}, \qquad (7)$$

where $N_{ord}$ represents the derivative order from 1-8, μ the mean, and $σ^2$ the variance.

Fig. 3 shows the scale to pseudo-frequency correspondence, described in (5), for the mother wavelets explored in this work. As several mother wavelets share $F_c$, as listed in Table II, these have been combined in Fig. 3. As the oscilloscope sampling rate, and thus Δ remained constant, the impact of $F_c$ on pseudo-frequency $F_a$ across

TABLE I. AVR INSTRUCTION SUBSET

| Instruction | Clock-Length | Type |
|---|---|---|
| sbi | 2 | bit / IO |
| nop | 1 | control |
| add | 1 | arithmetic |
| sub | 1 | arithmetic |
| cbi | 2 | bit / IO |
| push | 2 | transfer |
| pop | 2 | transfer |
| mul | 2 | arithmetic |
| eor | 1 | logic |
| movw | 1 | transfer |
| rjmp | 2 | program flow |

TABLE II. ASSESSED MOTHER WAVELET ATTRIBUTES

| Wavelet Family | Mother Wavelet | Symmetry | Center Frequency |
|---|---|---|---|
| Gaussian | gaus1 | anti-sym. | 0.2 Hz |
| | gaus2 | symmetric | 0.3 Hz |
| | gaus3 | anti-sym. | 0.4 Hz |
| | gaus4 | symmetric | 0.5 Hz |
| | gaus5 | anti-sym. | 0.5 Hz |
| | gaus6 | symmetric | 0.6 Hz |
| | gaus7 | anti-sym. | 0.6 Hz |
| | gaus8 | symmetric | 0.6 Hz |
| Mexican Hat | mexh | symmetric | 0.25 Hz |
| Morlet | morl | symmetric | 0.8125 Hz |

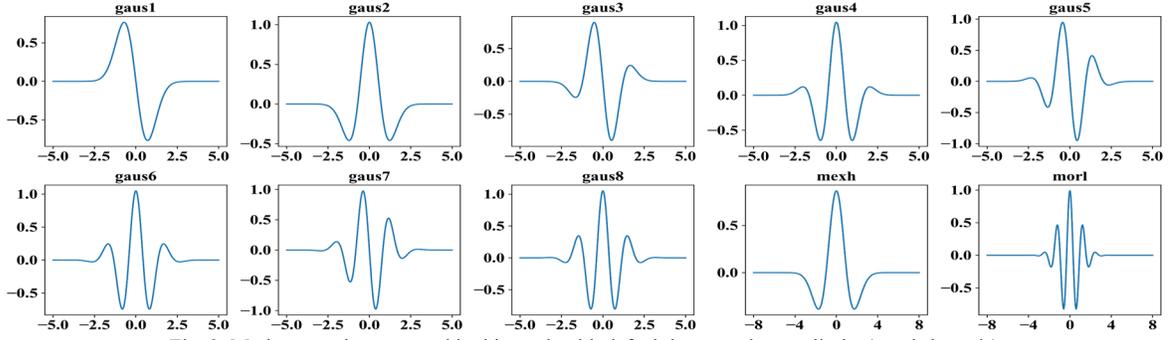
Fig. 2: Mother wavelets assessed in this work with default lower and upper limits (x-axis bounds)

scales becomes apparent. For instance, the *morl* wavelet provides coarser resolution, spanning approximately 200 MHz between scales 1-2, while *gaus1* spans approximately 50 MHz in the same range. However, with regard to bandwidth *morl*'s frequency range extends up to around 400 Mhz while *gaus1* tops-out around 100 MHz. Depending upon application and input signal, a higher bandwidth, such as with *morl*, or a finer resolution as with *gaus1*, may be desired.

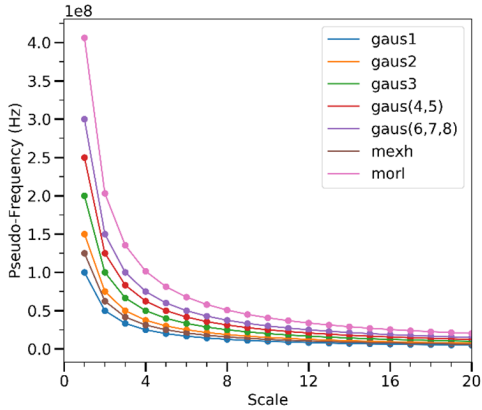
Fig. 3: Impact of wavelet scale on pseudo-frequency for assessed wavelets

### C. Wavelet Selection Overview

Since the application of varied wavelet functions to the same input will produce different results, selecting a suitable, or preferably optimal, mother wavelet for analysis is an important step. Some methods have leveraged wavelet function properties such as symmetry, orthogonality, and vanishing moments. Others explored the wavelet function in conjunction with the input data through methods such as visual similarity, cross-correlation, mean squared error, joint entropy measures, and more [23]-[26].

This work explored the use of similarity analysis through cross-correlation in the time-domain between ψ and a sample raw trace data file.

## IV. EXPERIMENTAL RESULTS

### A. Wavelet Accuracy

Each mother wavelet from Table II was assessed for impact on resultant classification accuracy. Five program loops from each trace file were analyzed. Wavelet coefficients for each scale from 1-50 were calculated. Fifteen trials of 100 training epochs for each mother wavelet were conducted, with test accuracy results shown in Fig. 4.

This experiment was conducted again with 20 trials at 125 training epochs and similar results were obtained: although it's relatively close, *gaus1* and *mexh* consistently achieved highest test accuracy, while *morl* was lowest in all cases.

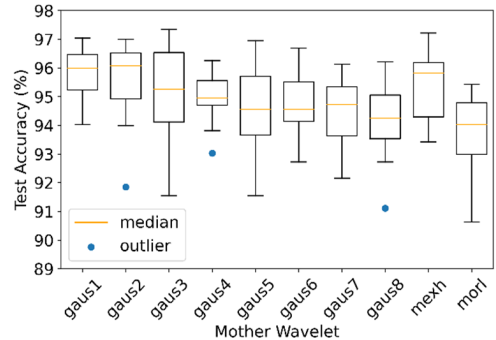
Fig. 4: Accuracy vs. mother wavelet with scales 1-50 over 15 trials

### B. Wavelet Time

#### 1) Coefficient Calculation Time

Although model training time is a primary factor in most efforts, with tens of thousands of clock-cycle samples, the time to calculate wavelet coefficients for each cannot be discounted. This is also an important consideration for integrity monitoring of a DUT to achieve near-real-time throughput. As part of the above wavelet accuracy experiments, wavelet coefficient calculation and time across all samples for each mother wavelet was also collected. Coefficients only need to be calculated once in preparation for CNN training, so this aspect was not repeated for each trial. Fig. 5 shows the coefficient calculation time for each mother wavelet, with *gaus* variants achieving similar times while *mexh* and *morl* were slightly longer in duration.

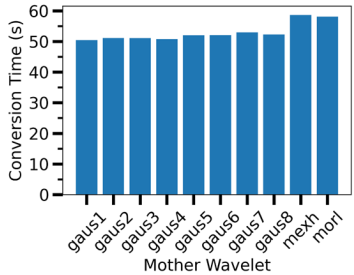

Fig.5: Example coefficient calculation time for 7,175 clock-cycle window samples (5 program loops from 5 files, 287 clock-cycles)

A dedicated experiment was then conducted, exploring the impact of scale as well as choice of wavelet on coefficient calculation time. In this experiment, the time to calculate wavelet coefficients for 10,000 clock-cycle window samples per wavelet at various scales between 1-1,000 was measured, and repeated for 100 trials. Fig. 6 shows the average time across trials for each mother wavelet at each scale. Three additional mother wavelets, *cmor*, *fbsp*, and *shan* were also included to demonstrate the significant impact both wavelet selection and scale can have in coefficient calculation times. In all cases there exists a linear relationship between scale and coefficient calculation time. Again, *mexh* and *morl* show similar times, slightly above (in general) that of *gaus* variants. However, at low scale ranges, *mexh* and *morl* appear slightly faster than *gaus* variants. Additionally, some separation is also shown between *gaus* variants, whereby higher orders account for longer durations. These factors can be observed more clearly in the inset of Fig. 6.

## 2) Training Time

The CNN training time (with a fixed number of epochs and no early-stopping) for each wavelet was also collected and assessed. No meaningful connection was established. Additional work would be required, to include early-stopping assessments, to identify if and how wavelet selection impacts CNN training time.

## C. Scale Accuracy

Most experiments in this work applied a fixed scale range covering a broad or focused range of interest relative to included pseudo-frequencies. In general, lower scales (higher frequencies) have proven more information dense and resulted in higher classification performance. This experiment explores the impact of scale on test set classification accuracy using sliding scale windows, conducted in two parts: scales 1 to 596 were explored with a window-width of 100, and sliding the window in increments of 5; scales 500 to 1000 were explored with a window-width of 100, and sliding the window in increments of 10. In both parts, significant overlap of windows was used to approximately localize performance changes. The *morl* wavelet, 40 epochs, and 10 trials were executed. The results are shown in Fig. 7.

As expected, the lower scale range (from 1 to around 140) offered the best performance for the *morl* wavelet. Classification continues to decrease as scales increase until plateauing and subsequently decreasing again. It is also worth noting that variance increases at higher scales.

## D. Colormaps

Wavelet transform coefficients are calculated and returned as a matrix **C** with dimensions $s \times t$, with $s$ representing the number of individual scales calculated, and $t$ the number of time samples. **C** can then be used to generate the input to the CNN, typically by applying a color-mapping whereby the wavelet transform coefficients are translated, for example, to Red Green Blue (RGB) values in accordance with a specified colormap, and dimensionality of the matrix now including three color-channels (in the RGB case).[1]

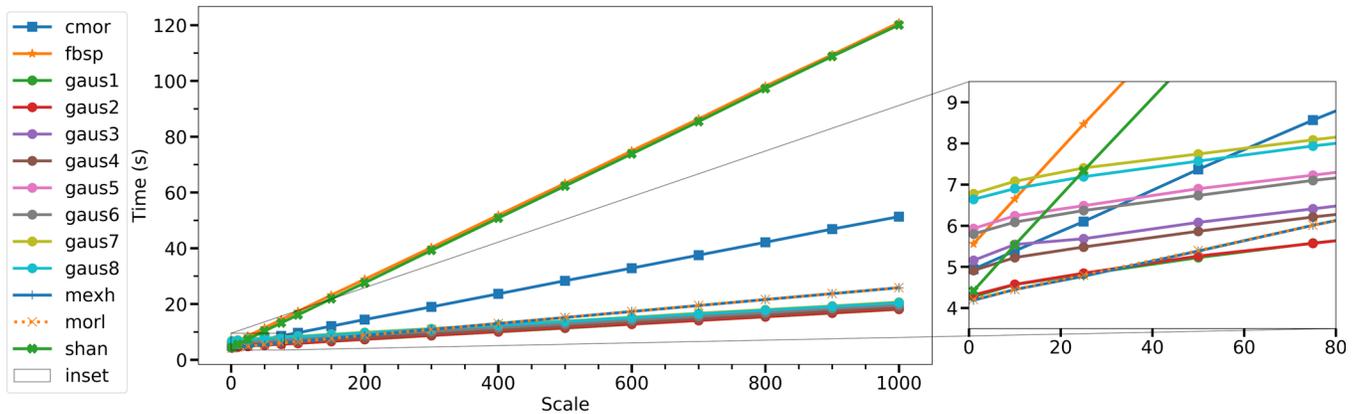

Fig. 6: Mother wavelet scale vs. time for 10k samples over 100 trials (average shown) with graph inset showing Gaussian separation by derivative order and time-duration crossover points of non-Gaussian mother wavelets

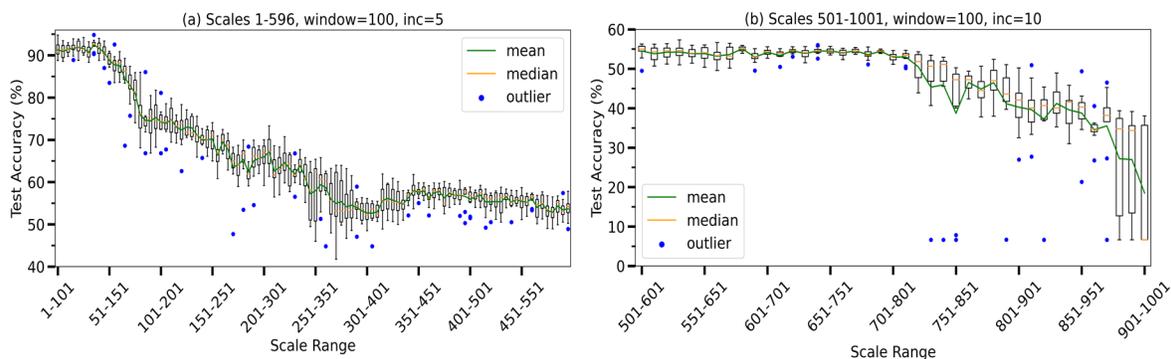

Fig. 7: Windowed scale vs. accuracy over 10 trials for *morl* mother wavelet, with decreasing performance approximately after scale window 40-140.

*1) Classification Accuracy*

The set of 176 colormaps from matplotlib [37] was used to assess the impact of colormap selection on resultant classification performance. This experiment was conducted for exploratory purposes to establish relative performance, and is an exception to the previously specified program loop and semi-optimized analysis set. It was performed on an earlier program loop consisting of the same instructions in Table I, but with different quantities and sequencing of each. Additionally, due to the amount of computation: less input data and a scaled-back CNN was used. Two trials were conducted using *morl* with wavelet coefficients from scale 1-251 and 100 epochs, for each colormap. The average classification accuracy of the test set across trials, grouped by colormap class is shown in Table III.

At the class level, there exists little difference between the top four classes, unless increased optimization is sought. The class outlier is *Qualitative*, which not only produced significantly lower performance but also higher variation within its membership. The top-2 by average accuracy was achieved with *cubehelix* and *twilight_r* colormaps of the *Miscellaneous* and *Cyclic* classes. However, the top-100 were all within about 1% of the maximum average accuracy achieved.

*2) Training Time*

Although the CNN training time (with a fixed number of epochs and no early-stopping) for each colormap was also collected and assessed, a meaningful connection was not established. Instead a general increase in training time was observed over the course of the evaluation. Additional work would be required, to include early-stopping assessments, to identify if and how choice of colormap impacts CNN training time.

*E. Wavelet Selection Example Results*

As mentioned in section III.C., a time-domain cross-correlation technique was explored for mother wavelet selection. Initial results included *gaus3* with the highest cross-correlation, and *morl* with the lowest. Further, many of the Guassian variants achieved similar but slightly lower cross-correlation to *gaus3*, while *mexh*'s cross-correlation was also near the top. The real outlier was *morl*, which resulted in significantly lower average and maximum cross-correlation values as shown in Table IV. Although these results do not completely match the primary wavelet analysis findings in this work, they appear fairly consistent, especially with regard to *morl* as a non-ideal choice.

TABLE III. COLORMAP CLASS ACCURACY

| Class | Number in Class | Class Avg. Accuracy | Class Std. Deviation |
|---|---|---|---|
| Cyclic | 6 | 92.1 % | 0.42 |
| Diverging | 28 | 92.1 % | 0.30 |
| Misc. | 32 | 91.7 % | 0.96 |
| Sequential | 86 | 91.5 % | 0.60 |
| Qualitative | 24 | 76.9 % | 3.16 |

TABLE IV. WAVELET CROSS-CORRELATION

| Mother Wavelet | Mean | Max |
|---|---|---|
| gaus1 | 0.168 | 0.801 |
| gaus2 | 0.200 | 0.839 |
| gaus3 | 0.216 | 0.946 |
| gaus4 | 0.217 | 0.900 |
| gaus5 | 0.207 | 0.929 |
| gaus6 | 0.190 | 0.822 |
| gaus7 | 0.168 | 0.807 |
| gaus8 | 0.145 | 0.706 |
| mexh | 0.202 | 0.861 |
| morl | 0.015 | 0.109 |

---

[1] CWT and/or scalogram subsampling, resizing, and other preparatory data munging techniques are vital but beyond the scope of descriptions in this paper.

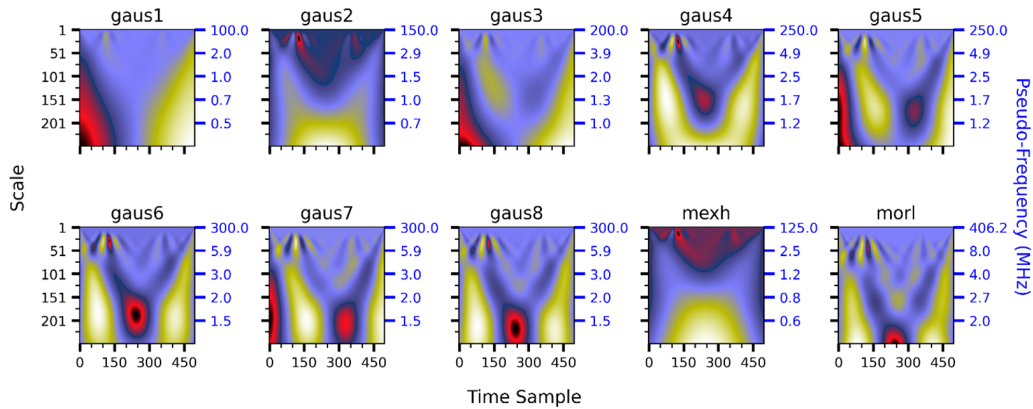

Fig.8: Example of same clock-cycle (program loop index 1) across mother wavelets at scales 1-250 with the *gist_stern* colormap applied.

## V. Conclusions

Wavelet selection impacts both classification performance and coefficient calculation time, as indicated by Figs. 4-6. Scales should be deliberately selected for a given wavelet and input signal under analysis based upon classification performance objectives and constraints to select the minimum quantity and lowest range of scales that achieve this result. As shown for *morl*, higher scales result in decreased classification performance (Fig. 7), and increased calculation time (Fig. 6). Correlation appears to be an average indicator of classification performance for wavelet selection, while selection based upon experimental results is more favorable.

The choice of colormap, when *Qualitative* is excluded, is not a significant factor in classification performance. Therefore, 1-channel grayscale was subsequently adopted. This provided sufficient classification performance, with lower memory requirements, and faster CNN training times as compared to 3-channel scalograms.

The *gaus1* mother wavelet was found to provide the best overall balance between classification performance and efficient coefficient calculation. Combined with 1-channel grayscale colormap to scalogram, a singular scale of 1 has provided greater than 99% classification accuracy in testing. Considering (6), a scale range of 1 to 22 may provide more robust results in practice.

This work demonstrates that wavelet selection and employment parameters have meaningful impact on analysis outcomes. Researchers should describe this information in their work to support reproducibility. Practitioners can use this information to make informed decisions and should consider optimizing these factors similarly to machine learning architecture and hyperparameters. Although these findings are only claimed for our application, it is believed they'll apply to other methodologies and use-cases, but additional work from the community is required to achieve generalized results.